\documentclass[12pt]{iopart}

\usepackage{iopams}
\usepackage{graphicx}
\usepackage{subfigure}
\usepackage{hyperref}
\usepackage{color}
\usepackage{url}
\usepackage[abs]{overpic}
\usepackage{epstopdf}
\usepackage[usenames,dvipsnames]{xcolor}
\usepackage{soul}

\begin{document}

\title[Phoresis in turbulent flows]{Phoresis in turbulent flows}

\author{Vishwanath Shukla$^{1,2}$, Romain Volk$^1$, Micka\"el Bourgoin$^1$ and Alain Pumir$^1$}
\address{$^1$ Laboratoire de Physique, ENS de Lyon, UMR CNRS 5672, Universit\'e de Lyon, France \\
$^2$ Service de Physique de l'Etat Condens\'e, Universit\'e Paris-Saclay, 
CEA Saclay, 91191 Gif-sur-Yvette, France}
\ead{research.vishwanath@gmail.com,
romain.volk@ens-lyon.fr,
mickael.bourgoin@ens-lyon.fr,
alain.pumir@ens-lyon.fr}
\date{\today}

\begin{abstract}
{Phoresis}, the {drift} of particles induced by scalar gradients 
in a flow, can result in an effective compressibility,  
bringing together 
or repelling particles from each other. Here, we ask whether this 
effect can affect the transport of particles in a turbulent flow. 
To this end, 
we study how the dispersion of a cloud of phoretic particles is 
{modified when injected in the flow, together with a blob of scalar,
whose effect is to transiently bring particles together, or push them away 
from the center of the blob. The resulting phoretic effect can be quantified 
by a single dimensionless number.}
Phenomenological considerations lead to simple predictions for the mean
separation between particles, which are consistent with results of 
direct numerical simulations. Using the numerical results presented 
here, as well as those from previous studies, we
discuss quantitatively the experimental consequences
of this work and the possible impact of such phoretic mechanisms in 
natural systems.
\end{abstract}

\noindent{\it keywords}: turbulence; turbulent transport; phoresis; direct numerical simulations


\maketitle

\section{Introduction} 

The transport of particles and macro-molecules 
in a flow can be strongly affected by a scalar quantity present in the fluid.
The phenomenon of phoresis investigated here results from a drift velocity
of the particles, $\mathbf{v}_d$, proportional
to the gradient of a scalar quantity, $\theta(\mathbf{x}, t)$ :
$\mathbf{v}_{d} = \alpha\nabla\theta(\mathbf{x},t)$.
The scalar field involved could be temperature (thermophoresis), a chemical 
species (chemophoresis), or salt concentration (diffusiophoresis). 
The magnitude and sign of the phoretic mobility coefficient $\alpha$ 
depends on the nature of the interaction of the particle with the scalar 
field~\cite{anderson1989colloid}. Electrophoresis and 
magnetophoresis~\cite{li2008encyclopedia,zborowski2015magnetophoresis}, 
induced by a drift proportional to the electric or magnetic field, respectively, 
provide two extra examples of phoresis with potential practical utility.

It is well-known that the diffusive transport of particles 
and of macro-molecules in a laminar flow is generally very slow because of
their very small intrinsic molecular diffusion coefficient.
Recent studies show that a significant enhancement of the transport
properties can be achieved
by addition of a small salt concentration to the solution,
which induces a phoretic mobility in response to the salt gradients
~\cite{bib:NatAbecassis2008,abecassis2009NJP}. 
The enhanced mobility of the colloids observed in microfluidic channel 
experiments with salt gradients~\cite{bib:NatAbecassis2008,abecassis2009NJP} 
was initially interpreted in terms of an effective diffusion 
aided by diffusiophoresis. The inferred diffusion coefficient is larger than 
what is expected from a simple Brownian motion of colloidal particles and for
certain salts it is almost two orders of magnitude larger. This makes 
diffusiophoresis an important concept with a practical utility, which can be 
used to selectively control the mobility of colloids or macro-molecules in 
microfluidic devices for scientific and/or industrial purposes. 
Similar arguments based on effective diffusivity were later used to
explain the experimental results on enhanced or delayed mixing of colloids in 
chaotic flows~\cite{deseigne2014softmat}.

However, more recent studies~\cite{PREVolk2014,Mauger16} showed that a 
description of these phenomena in terms of a {modulated diffusion coefficient
does not capture all the physics at play}. In particular, it fails to {explain} the 
evolution in space and time of the colloids concentration, especially
mixing and de-mixing at short times. 

These studies stressed the importance of the effective \textit{compressibility} 
of the velocity field $\mathbf{v}$ seen by the particles:  
$\nabla\cdot\mathbf{v}\neq 0$. Numerical simulations  and experiments 
employing diffusiophoresis in chaotic flows have conclusively shown 
that the compressibility strongly affects the transport of particles. 
In particular, it was found that the compressible nature of the velocity 
field acts as a source of colloids concentration variance ~\cite{PREVolk2014}. 
As a consequence the whole mixing process is modified by the phoretic 
transport, leading to either mixing or de-mixing depending on the local 
environment. Moreover recent experiments in a chaotic flow clearly demonstrated 
that diffusiophoresis modifies the properties of the particle distribution 
not only at small-scales, but even at the largest scales of the 
mixing process~\cite{Mauger16}.

Here, we ask how the transport of colloids by a turbulent flow is affected 
by phoresis. Turbulent motion in fluids involves a wide range of length and 
time scales and is known to enhance mixing by generating strong gradients 
of advected scalar fields. The study of the transport of particulate matter 
in turbulent flows is an important problem, with far reaching 
implications for a wide spectrum of applications, such as
rain initiation in warm clouds~\cite{Shaw03} or 
controlling industrial flows~\cite{BalEaton:2010}. 
The velocity field experienced by inertial particles, 
with a finite-size and/or different density compared to the carrier fluid, 
is compressible~\cite{MaxRil83,Gatignol83}. It is known that this effective 
compressibility results in an inhomogeneous distribution of inertial 
particles~\cite{BFF01,FFS02,FalkPum04} in contrast to the homogeneous 
distribution of pure tracer particles in turbulent flows.
Therefore, it is natural to use the concepts and current understandings of 
the role played by the effective compressibility in turbulent transport 
processes, when extended to phoretic particles driven by turbulent 
scalar field~\cite{Falkovich2014,Schmidt2016,de2016clustering,mitra2016turbophoresis}.

In the following,
we examine the consequences of the turbulent fluctuations on the
dynamics of a cloud of particles undergoing phoresis. In the spirit
of the original experiments, documenting phoretic effects by injecting 
locally colloids and salt~\cite{bib:NatAbecassis2008,abecassis2009NJP},
we focus on the effect of phoretic compressibility at short times, 
after releasing a cloud of salt and colloids, of characteristic size 
$\ell_c$, in a turbulent flow. Although it is transported by a turbulent 
flow, such a cloud of size $\ell_c$ will grow ballistically during 
a characteristic time $t_c$, which can be estimated using the standard 
phenomenological description of turbulence (see Section 3, and ref.~\cite{Frisch}).

Dimensionally, the effective compressibility $( \nabla \cdot \mathbf{v} ) $, 
where $\mathbf{v}$ is the velocity seen by the particles, is an inverse 
time scale, which should be compared to the typical time scale of the motion;
note that $\mathbf{v}$ includes both the velocity of the carrier flow $\mathbf u$
and the phoretic drift $\mathbf{v}_d$. 
Therefore, we introduce a dimensionless number,
$\Phi_{\mathcal{C}} \sim ( \nabla \cdot \mathbf{v} ) \times t_c$, which characterizes
the competition between turbulent dispersion and compressibility. We use 
these phenomenological arguments, which are corroborated by Direct Numerical 
Simulations (DNS) of the Navier-Stokes equations (NSE), to show that $\Phi_{\mathcal{C}}$
is the parameter which controls the mixing of the phoretic particles
in this problem.
In particular, we show that the maximum contraction of the particle 
cloud occurs in the ballistic regime of turbulent dispersion, and
varies as $\exp(-\gamma\Phi_{\mathcal{C}})$. Our results show that 
$\gamma$ has similar values for different Reynolds numbers
and particle cloud sizes, thereby suggesting an universal behavior.


Phoretic effects have also been observed to play a role in randomly stirred
chaotic flows in two dimensions (2D)~\cite{PREVolk2014}, in a
regime where the scalar field is in a statistically stationary state.
We discuss the expression of $\Phi_\mathcal{C}$ in such cases.
Our estimates, obtained by using realistic values of the physical parameters,
suggest that phoretic 
effects should also be observable in turbulent flows in steady 
state configuration at very small scales.

\begin{figure}[tb]
\begin{center}
\resizebox{0.75\linewidth}{!}{
\includegraphics[scale=0.5]{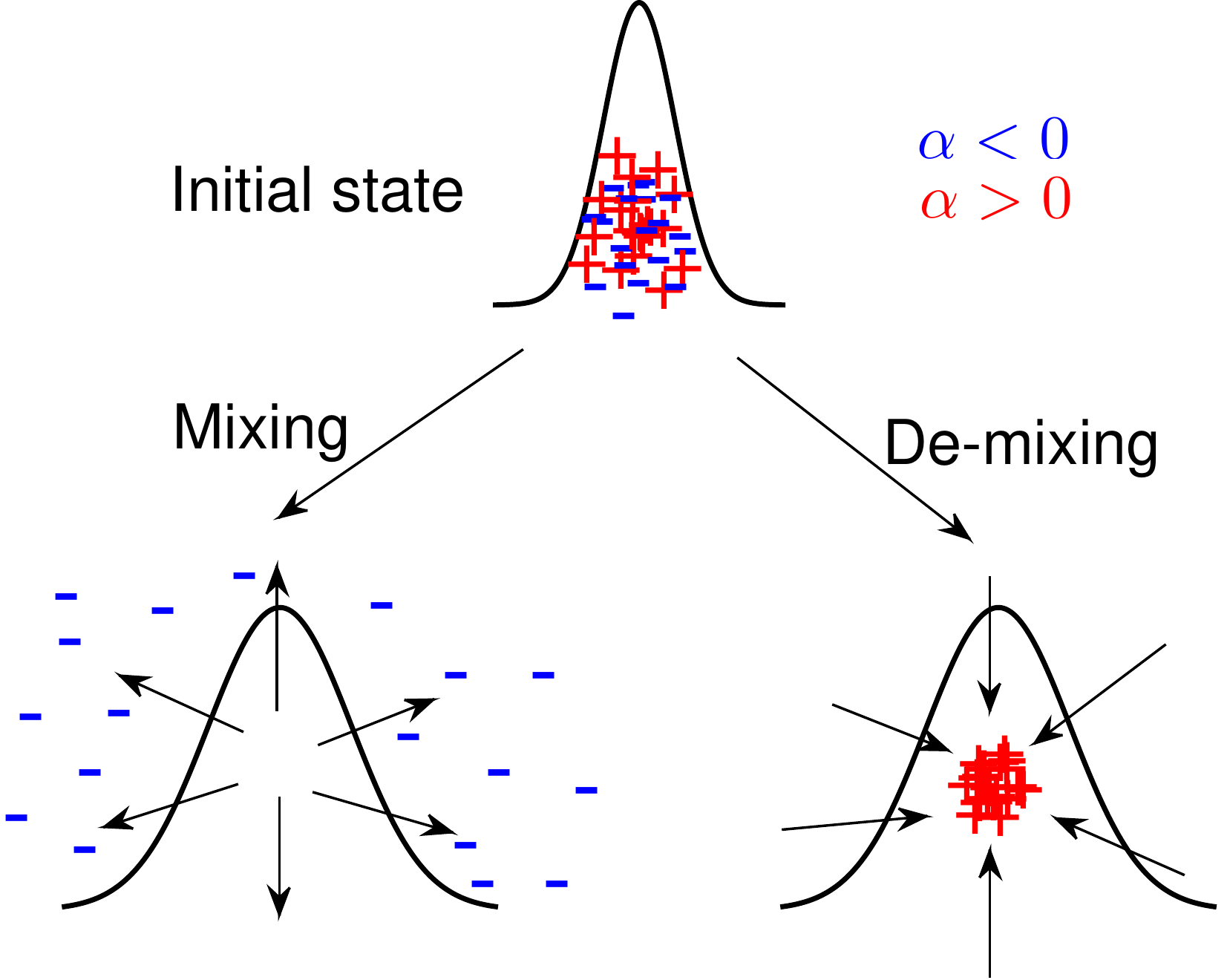}
}
\end{center}
\caption{{\bf Turbulent phoresis.}
Schematic diagram showing the initial configuration of the scalar field
and the resulting phoretic motion of the colloids. 
Particles with positive ($\alpha>0$, red ``+'' symbol) and negative 
($\alpha<0$, blue ``-'' symbol) phoretic mobilities
are  initially distributed in a Gaussian blob of scalar field (upper narrow Gaussian
profile). Scalar mixing in a turbulent fluid is indicated by drawing a broader scalar profile; 
depending on the sign of the phoretic mobility $\alpha$, the particles either tend to cluster
(de-mixing) or separate from each other (mixing).
}
\label{fig:sketch}
\end{figure}

\section{Methods} 

Specifically, we consider here a simple problem, whereby a blob
of scalar field with a simple Gaussian profile is released
in a turbulent fluid, along with a cloud of particles, as illustrated
in \fref{fig:sketch}.
The scalar field decays, as it is mixed by the turbulence.
During this process, strong scalar
gradients develop and this drives the phoretic motion
of the particles. We stress that this configuration is experimentally 
realizable, for example in a turbulent jet or a water-tunnel, where the
statistical properties of the cloud of particles and their 
dispersion could be accurately measured. 
We assume here that the flow is incompressible, so that its description is based on the 
(NSE)
\begin{equation} \label{eq:NSeq}
\frac{\partial \mathbf{u}}{\partial t} + \left(\mathbf{u}\cdot\nabla\right)\mathbf{u}
= -\nabla p + \nu\nabla^2\mathbf{u} + \mathbf{f}
\end{equation}
where $\nabla\cdot\mathbf{u}=0$ imposes the incompressibility condition on the three-dimensional
 (3D) velocity field $\mathbf{u}$. $p$ is the pressure field, $\nu$ is the 
fluid viscosity and $\mathbf{f}$ is the forcing term, acting at large scales,
which maintains the turbulent flow statistically stationary.
The spatio-temporal evolution of the scalar field fluctuations $\theta(\mathbf{x},t)$
is governed by the advection-diffusion equation
\begin{equation} \label{eq:addif}
	\frac{\partial \theta}{\partial t} + (\mathbf{u}\cdot\nabla)\theta = 
	\kappa\nabla^2\theta,
\end{equation}
where $\kappa$ is the diffusion coefficient of the scalar field constituents,
whose spatial average $\left<\theta\right>_{spatial}$ is zero.
We furthermore assume that the molecular diffusion coefficient of the 
colloidal particles of interest here
is orders of magnitude smaller than the constituents of the scalar field. Therefore,
we are interested in the limiting case where the particles are advected by the
combined velocity field of the turbulent flow $\mathbf{u}$ and the phoretic 
contribution $\mathbf{v}_d=\alpha\nabla\theta$, $\alpha$ being the phoretic mobility
coefficient, quantifying the efficiency of the gradient to generate an additional 
drift motion for the particles.
To study the dynamics of this assembly of discrete particles
undergoing phoresis in a turbulent environment, we adopt a Lagrangian framework,
in which the velocity of a particle, labelled by the index $i$, and 
located at position $\mathbf{X}_i(t)$ is given by 
\begin{equation}\label{eq:partvel}
	\mathbf{v}_{i} = \mathbf{u}(\mathbf{X}_{i},t) + \alpha\nabla \theta(\mathbf{X}_{i},t),
\end{equation}
where $d\mathbf{X}_{i}/dt= \mathbf{v}_{i}$.

As briefly sketched in the previous section, the phenomenological 
approach of particle dynamics used here
is based on
the standard (Kolmogorov-Obukhov) phenomenological description of
turbulence~\cite{Frisch}.
A three-dimensional (3D) turbulent flow is sustained by injecting energy 
at a rate $\varepsilon$ per unit mass, 
at the large length scales, 
$\ell\sim L_0$, comparable to the system size. 
This energy is transferred to smaller length scales, down 
to the scale $\eta$ where energy is dissipated by viscosity: 
$\eta = (\nu^3/\varepsilon)^{1/4}$. The range of scales, defined by
$\eta \ll \ell \ll L_0$, is defined as the inertial scales. 
The transfer of energy is a result of the 
nonlinear interactions, represented by the advection term in the 
NSE \eref{eq:NSeq}.
An (almost) self-similar structure is often postulated over the inertial 
range of scales, which is characterized by a constant flux of energy, $\varepsilon$~\cite{Frisch}.

We test our phenomenological arguments and make them more quantitative by 
numerically determining the statistical properties of this system.
To this end, we perform
DNSs of the NSE \eref{eq:NSeq} and the advection-diffusion 
equations \eref{eq:addif}
in 3D to determine the fluid velocity field $\mathbf{u}$ and 
the scalar field $\theta$, respectively;
we then use them to numerically determine the trajectories of the particles.
We solve the NSE in a triply-periodic domain by
using the pseudospectral code, as described, e.g. in Refs.~\cite{Pum94,FalkPum04}.
The flow is maintained statistically stationary by
forcing the
velocity field at large scales (small wave numbers $k$)~\cite{Pum94}. 
Simulations were carried out at moderate resolutions, with $ 96^3$, $192^3$
and $384^3$ Fourier modes. Adequate spatial resolution for the 
velocity field was imposed, by
maintaining the product $k_{max} \eta \ge 1.5$. This allowed us to simulate 
flows up to a Taylor-scale Reynolds number $R_\lambda = 175$. To address the
more demanding resolution criteria for the scalar field~\cite{Pum94}, we work
here with a dimensionless ratio 
$\nu/\kappa$ equal to $1/2$. 
We refer here to this dimensionless ratio as the Prandtl number 
$Pr$,
a terminology which is generally used when for a scalar field (the 
corresponding dimensionless ratio is known as the Schmidt number, $Sc$, in the 
case of a concentration field).

Here, we start with an initial configuration, where we release
in a turbulent flow a Gaussian blob of scalar:  
\begin{equation}\label{eq:initscl}
\theta(\mathbf{x},t=0) = \theta_0\exp\left(-\frac{\mathbf{x}^2}{2\ell^2_c}\right)
\end{equation}
containing $2000$ particles which are randomly distributed within it with the same Gaussian distribution. The size $\ell_c$ then serves as a convenient
measure both of the scalar blob and the particle cloud.
The effect of the scalar is to attract particles close to the center when 
$\alpha \theta_0 > 0$, and repel them away from the center when $\alpha \theta_0 < 0 $, 
as illustrated in \fref{fig:sketch}.

A list of the simulations carried out in this work is provided in 
Table~\ref{table:param} and Table~\ref{table:Phic}, see~\ref{appendix:supmat}.

\section{Results} 
\label{sec:results}

\subsection{Competition between phoretic effects and turbulence mixing} 

As already stressed, the velocity field seen by the particles, 
\eref{eq:partvel}, is compressible
and its divergence is proportional to the Laplacian of the scalar field
$\nabla\cdot\mathbf{v}=\alpha\nabla^2\theta$. The divergence field
$\mathcal{C}(\mathbf{x},t)\equiv\nabla\cdot\mathbf{v}$ can serve as a local 
indicator of the compressibility. With \eref{eq:initscl} as the initial 
condition for the scalar field, the initial compressibility is given by 
$\mathcal{C}_0 \sim \alpha \theta_0 /\ell_c^2$. 
Starting from initial condition, the blob of fluid which contains both scalar 
and particles will be advected by turbulent motions. Particles will 
then experience compressible effects at the scale $\ell_c$ (of the order 
$\mathcal{C}_0$), until the scalar field is distorted, 
which is achieved in a time $t_c$ corresponding to the eddy turnover 
time at scale $\ell_c$. Such a time scale can be estimated following 
standard phenomenology of turbulent flows~\cite{Frisch}. As $\ell_c$ 
lies in the inertial range, $t_c$ does not depend on the viscosity and 
reads $t_c \sim (\ell_c^2/\varepsilon)^{1/3}$, where $\varepsilon$ is 
the energy injected per unit mass. This leads us to define the 
dimensionless parameter that characterizes the competition between turbulent
dispersion and compressibility effects at a given scale $\ell_c$ by:
\begin{equation} \label{eq:phic}
\Phi_{\mathcal{C}}=\alpha\frac{\theta_0}{\ell^2_c}t_c.
\end{equation}
We chose to define $\Phi_{\mathcal{C}}>0$ for the attracting case which 
corresponds to $\alpha \theta_0>0$, and fixed $\theta_0=1$ in all our DNS runs.
Note that $t_c$ is Kolmogorov-Obukhov phenomenological estimate of the 
time for which an eddy of the size of the scalar blob
$\ell_c$ survives in a turbulent flow.

\begin{figure*}
\begin{center}
\resizebox{0.9\linewidth}{!}{
\includegraphics[scale=0.45]{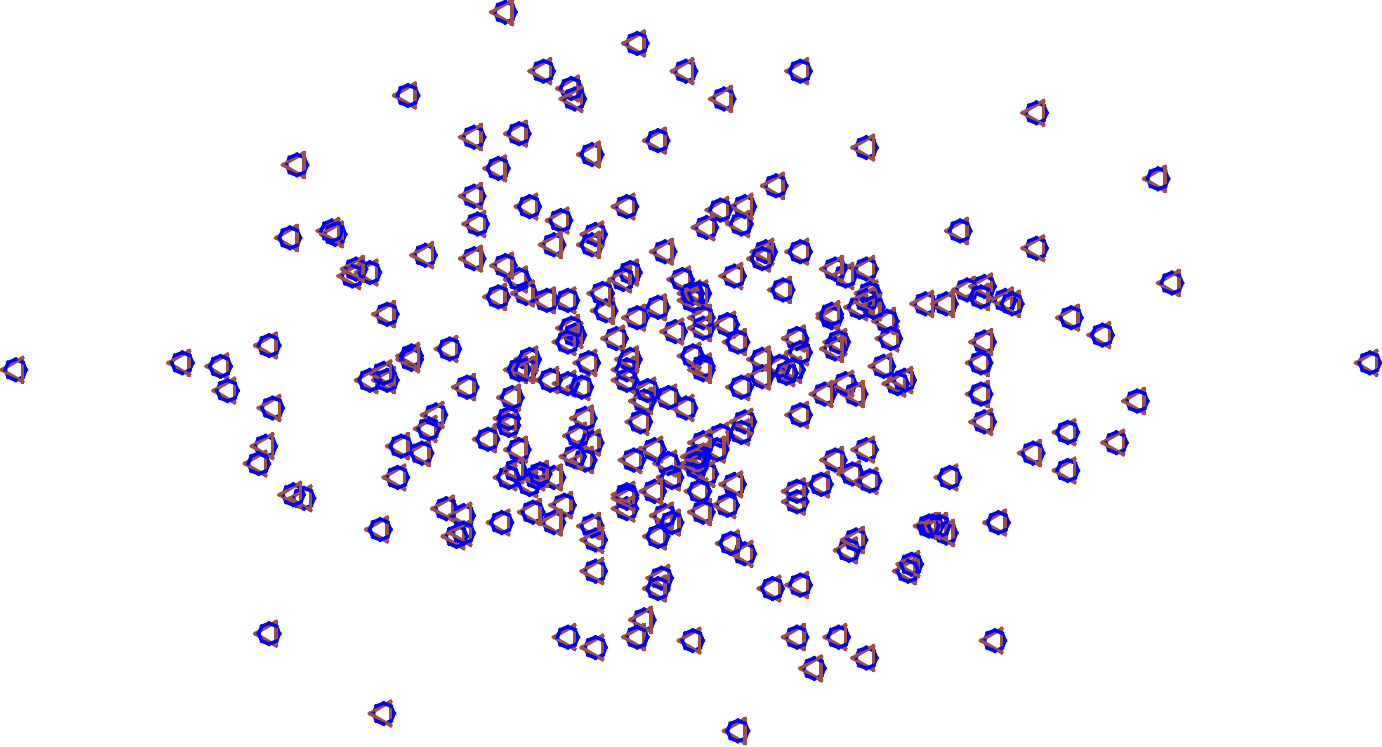}
\put(-130,-12){\large{{\bf(a)} $t/t_c=0$}}
\includegraphics[scale=0.45]{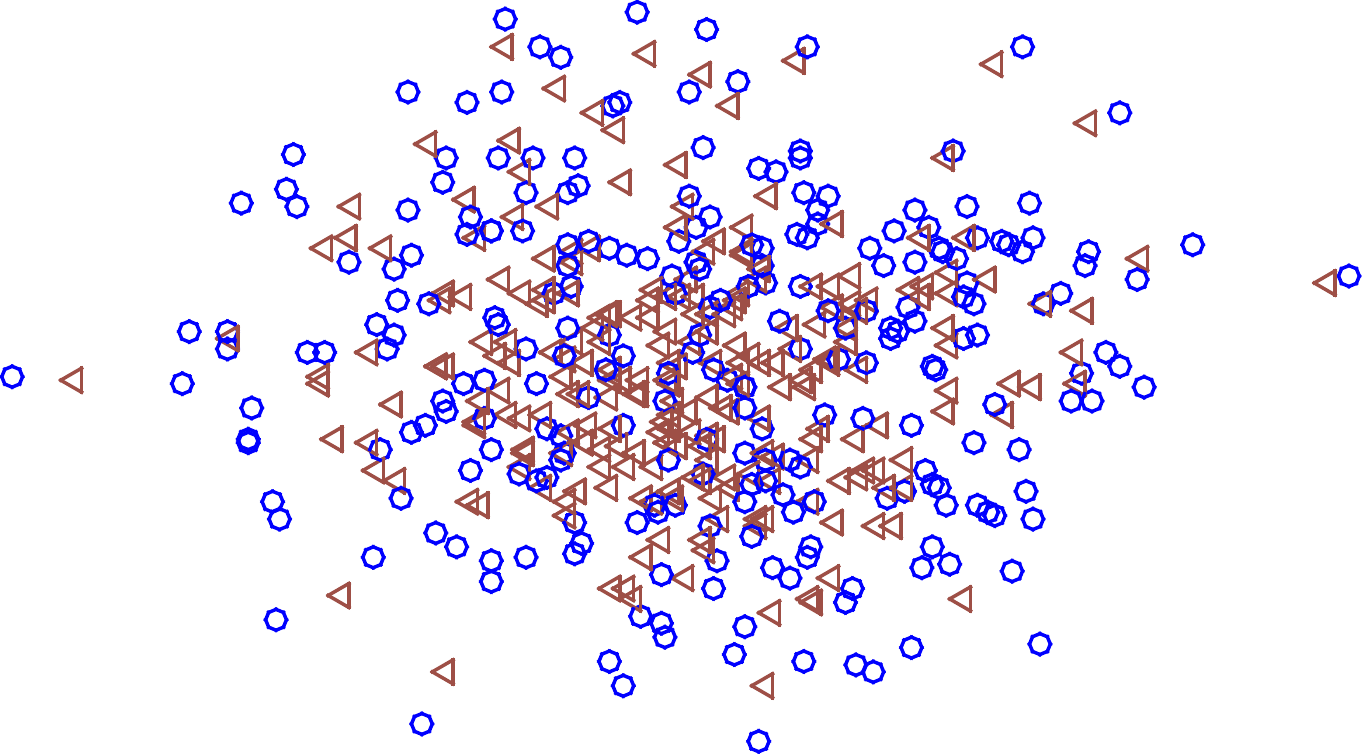}
\put(-120,-12){\large{{\bf(b)} $t/t_c=0.07$}}
\includegraphics[scale=0.45]{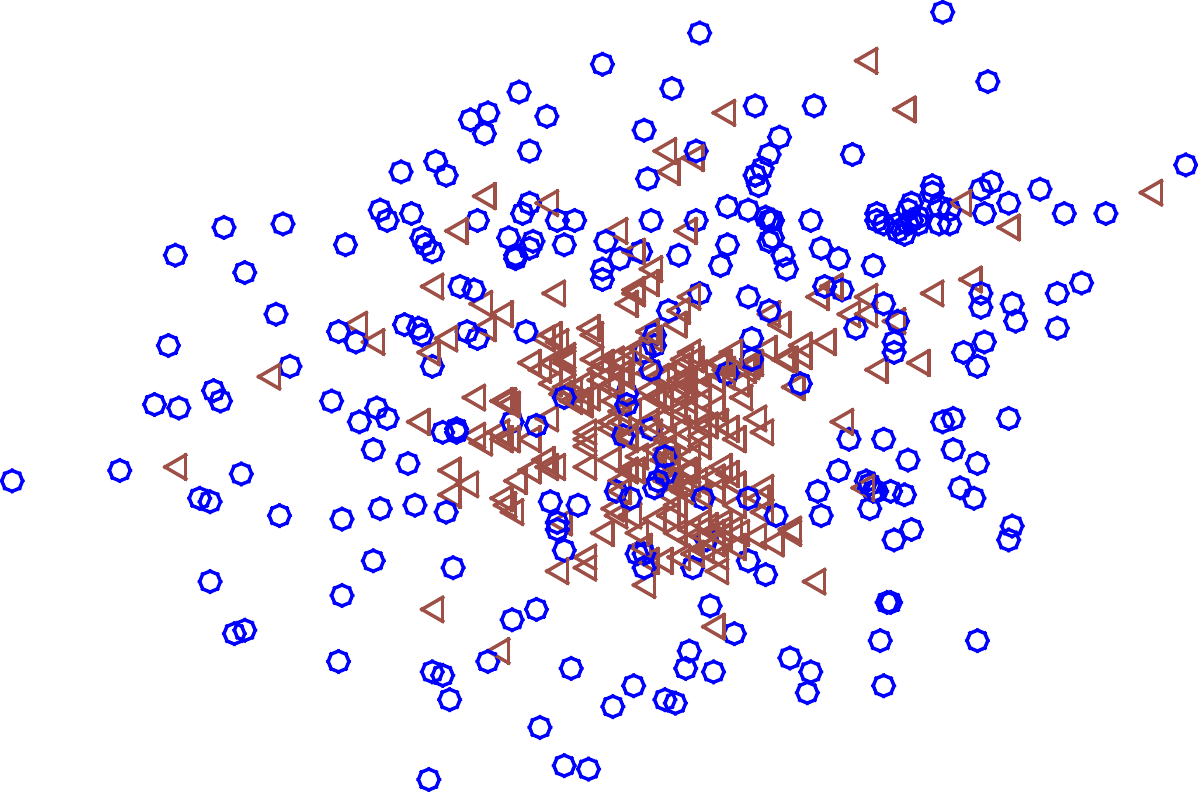}
\put(-110,-12){\large{{\bf(c)} $t/t_c=0.22$}}
}
\end{center}
\caption{
{\bf Mixing/de-mixing of phoretic particles.} Positions of two particle types 
at three different times: $t/t_c=0$ (a), $t/tc=0.07$ (b), and $t/t_c=0.22$ (c). 
The particles correspond to two different values of the dimensionless number 
$\Phi_{\mathcal{C}}=-5.81$ (blue circles) and $\Phi_{\mathcal{C}}=5.81$ 
(brown triangles) to same local environment, 
from the DNS run with $R_{\lambda}=95$ and $\ell_c/\eta=20.8$. 
The particles with $\Phi_{\mathcal{C}}>0$ de-mix, whereas the ones 
with $\Phi_{\mathcal{C}}<0$ undergo enhanced mixing, for a scalar blob with 
the Gaussian profile in a turbulent flow.}
\label{fig:partcloud}
\end{figure*}

\subsection{Mixing and de-mixing of a turbulent cloud} 

In \fref{fig:partcloud} we show the positions of two particle types at 
three different times: (a) $t/t_c=0$, (b) $t/tc=0.07$, and (c) $t/t_c=0.22$. 
These two particle types correspond to two different values of 
$\Phi_{\mathcal{C}}=5.81$ (brown triangles, $\alpha>0$) and 
$\Phi_{\mathcal{C}}=-5.81$ (blue circles, $\alpha<0$).
Starting at identical initial positions at $t=0$ (left), we observe that for 
$\Phi_{\mathcal{C}}>0$ (brown triangles, $\alpha>0$) the particle cloud shrinks 
as a whole, whereas for $\Phi_{\mathcal{C}}<0$ (blue circles, $\alpha<0$) the 
particles in the cloud move away from each other, resulting in an overall expansion.

\begin{figure*}[tb]
\begin{center}
\includegraphics[scale=0.45]{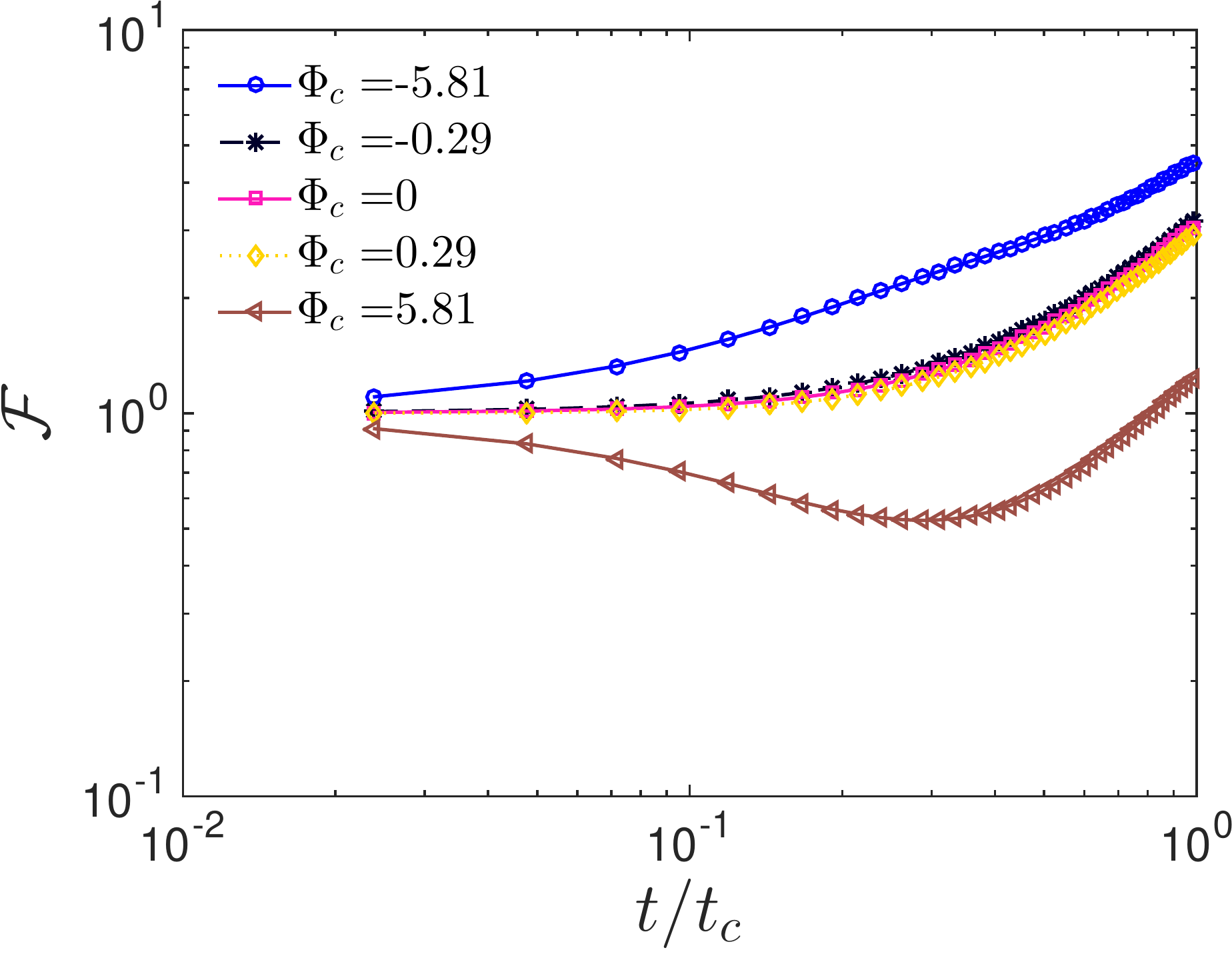}
\put(-40,40){\large{\bf(a)}}
\includegraphics[scale=0.45]{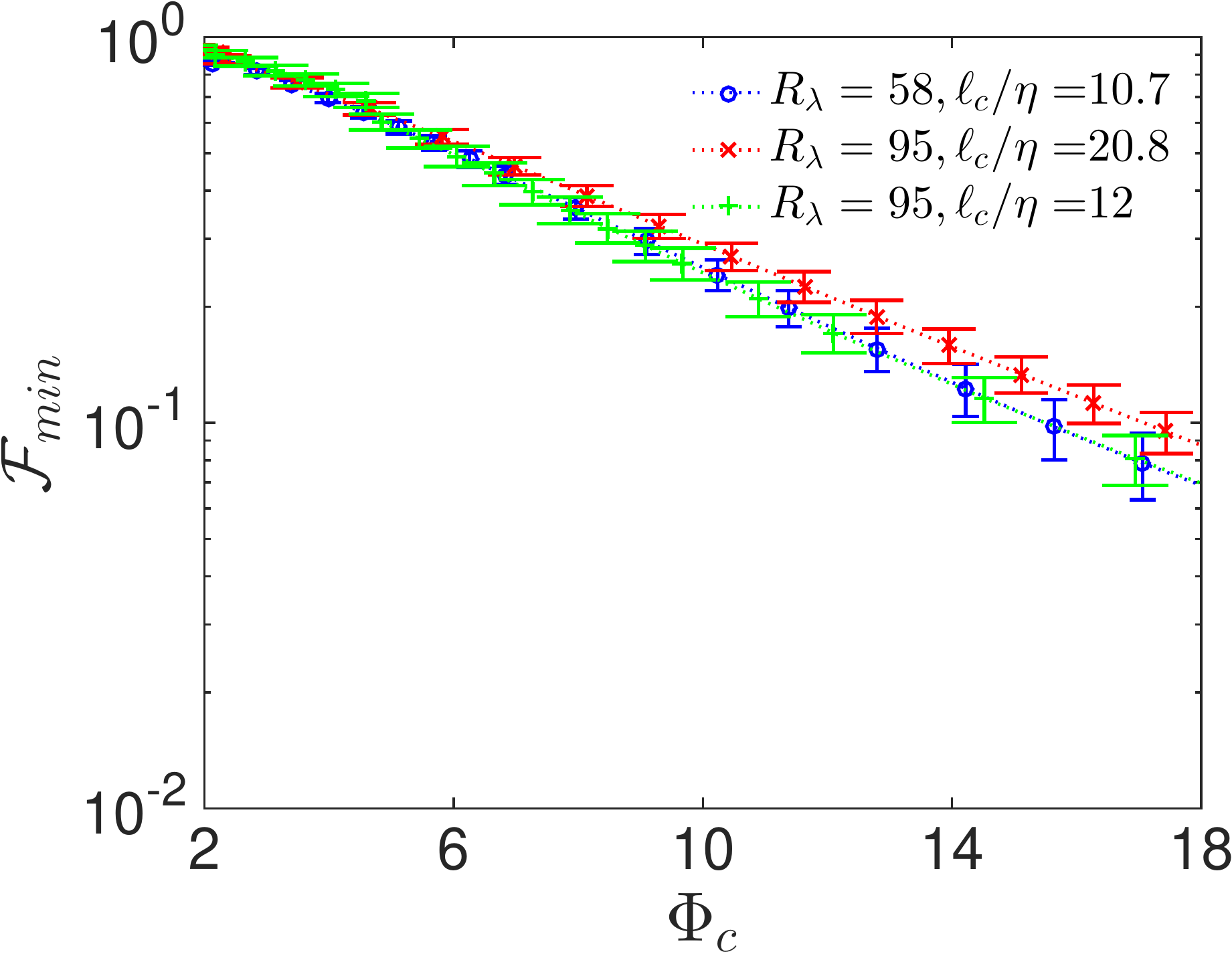}
\put(-190,33){\includegraphics[scale=0.21]{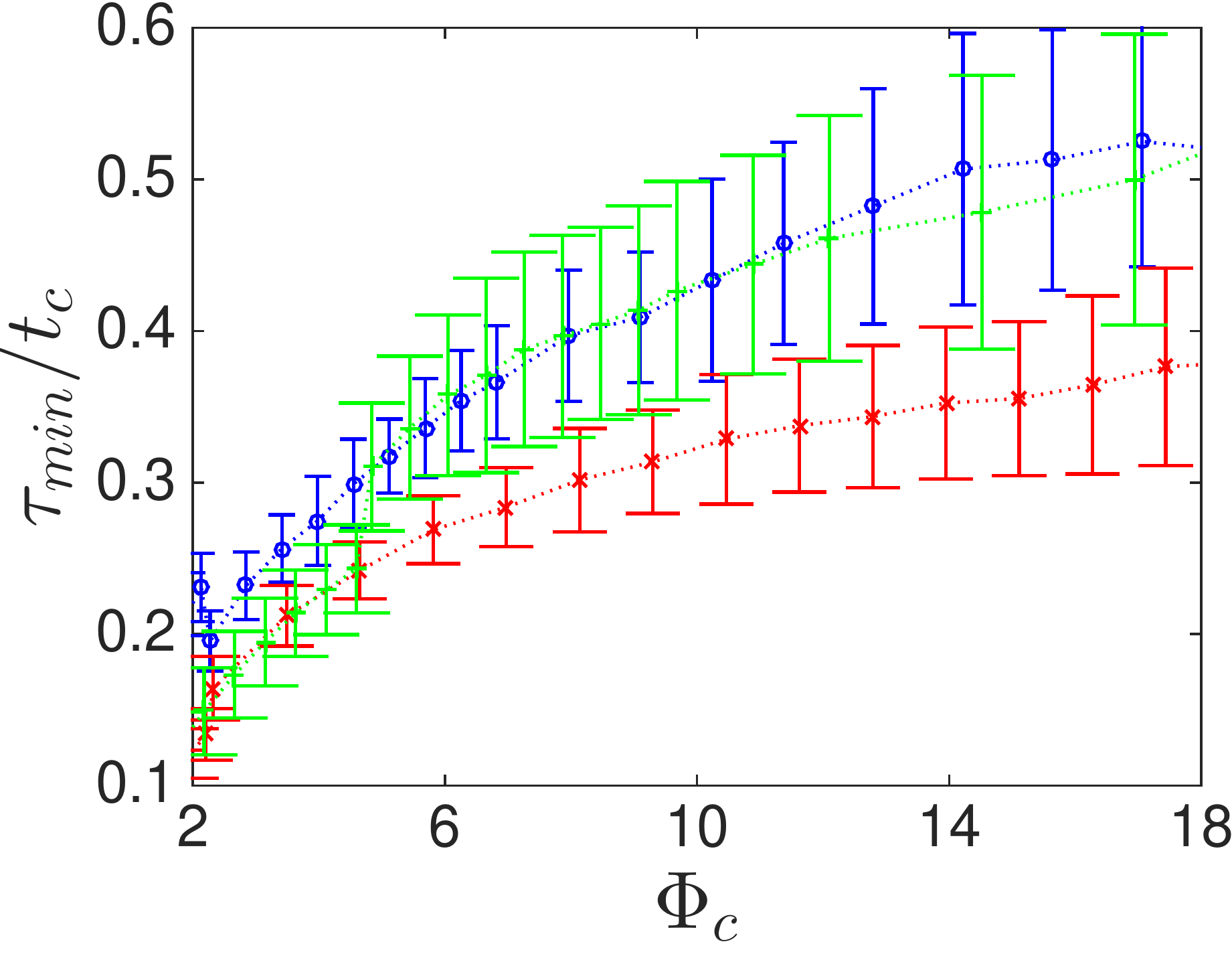}}
\put(-40,40){\large{\bf(b)}}
\end{center}
\caption{
{\bf Quantifying the de-mixing.}
(a) Temporal evolution of the ratio of mean relative separation to its value at $t=0$,
$\mathcal{F}(t) \equiv r^2(t)/r^2(t=0)$, for different values of the dimensionless
number $\Phi_{\mathcal{C}}$ characterizing the effective compressibility,
from the DNS run with $R_{\lambda}=95$ and $\ell_c/\eta=20.8$; 
$t_c$ is the time scale for which an eddy of size $\ell_c$ survives in a turbulent flow.
(b) Log-linear plots of minimum of $\mathcal{F}$ versus $\Phi_{\mathcal{C}}$. 
$\mathcal{F}_{\rm min}$ is representative of the maximal de-mixing attained.
Inset: Plots of $\tau_{\rm min}$ versus $\Phi_{\mathcal{C}}$, shows that the
time to achieve maximum de-mixing (or contraction of the particle cloud) initially 
increases and then saturates at large values of $\Phi_{\mathcal{C}}$.
}
\label{fig:r2ncomp}
\end{figure*}

To quantify the coherent expansion or the contraction of the particle cloud at short times,
we measure the mean relative separation between the particles $r(t)$ and its ratio to its 
value at $t=0$
\begin{equation}\label{eq:ratior}
	\mathcal{F}(t)\equiv\frac{r^2(t)}
	{r^2(t=0)},
\end{equation}
where ${r}^2(t)= \langle (\mathbf{X}_i(t)-\mathbf{X}_j(t))^2 \rangle_{1\leq i<j \leq N}$. 
In order to get better statistics, $8$ blobs of scalar containing particles are released 
at the same time with distances larger than half the box size. The simulation is then 
run and stopped before the different deformed blobs overlap.
In \fref{fig:r2ncomp} (a) we show the temporal evolution of $\mathcal{F}$ for
five different values of $\Phi_{\mathcal{C}}$ for the DNS run for which 
$\ell_c/\eta=20.8$ and $R_{\lambda}=95$. The pink curve with squares represents
the case of a pure tracer $\Phi_{\mathcal{C}}=0$ ($\mathcal{C}=0$). For the phoretic
particles with $\Phi_{\mathcal{C}}<0$ (blue curves with circles and black curves with stars) 
$\mathcal{F}$ increases faster than it does for the pure tracers. This represents a case where the
local environment selectively assists in the faster dispersal of these particles, thereby
resulting in an enhanced mixing or expansion of the particle cloud at short times.
In a direct contrast to this, for phoretic particles with 
$\Phi_{\mathcal{C}}>0$ (yellow curve with diamonds and brown curve with 
triangles),
$\mathcal{F}$ first decreases until it reaches a minimum value and then it rises rapidly.
This clearly represents a case of de-mixing aided by the local environment, whose
extent depends on the value of $\Phi_{\mathcal{C}}$ and has been quantified in terms
of the coherent contraction of the particle cloud monitored by measuring $\mathcal{F}$.

In the following, we focus on the case $\Phi_{\mathcal{C}} > 0$, which results in de-mixing 
(see \fref{fig:r2ncomp}).
We note that the lower the value reached by $\mathcal{F}$, the stronger 
is the de-mixing.
We propose here a simple description of this effect, based on the standard
Kolmogorov-Obukhov phenomenological theory of turbulence~\cite{Frisch},
resulting in a prediction of $\mathcal{F}_{\rm min}$ as a function of 
$\Phi_{\mathcal{C}}$.  

We can use \eref{eq:partvel} and \eref{eq:initscl} to write 
\begin{equation}\label{eq:pvelincre}
	\frac{d \delta \mathbf{X}}{d t} = \delta  \mathbf{u} +  \alpha \delta {\nabla} \theta,
\end{equation}
where $\delta \mathbf{X}$ (resp. $\delta \mathbf{u}$) is the difference between 
the positions (resp. velocities) of two phoretic particles (indices $i,j$ have been 
dropped for the sake of simplicity). 
Note that the first term on the right hand side 
(RHS) in \eref{eq:pvelincre} is a velocity increment.
In the absence of phoretic effect ($\alpha = 0$), the term $\delta \mathbf{u} $
is responsible for turbulent particle dispersion. 
This term averages to zero when the colloids are released,
as the particle positions are {\it a-priori} not correlated with the flow 
velocity field~\cite{Bragg+14}.
At small times, this term 
contributes only a small correction to the constant term, 
in particular when $| \Phi_{\cal C} | \gtrsim 1$,
see \fref{fig:r2ncomp}.
The second term is the phoretic contribution which produces the effect shown in 
\fref{fig:partcloud} and \ref{fig:r2ncomp}. At a qualitative level, this term 
is smooth and nearly isotropic so that one has ($t <t_c$, 
$\delta \mathbf{X} \sim \delta \mathbf{X}_0$)
\begin{equation}\label{eq:pvelincre2}
\frac{d \langle \delta \mathbf{X}^2 \rangle}{d t} \propto 
- \alpha \frac{\theta_0}{\ell^2_c} \langle \delta \mathbf{X}^2 \rangle,
\end{equation}
when averaging over particle pairs in the ballistic regime of turbulent dispersion.
As long as the injected blob of scalar keeps its identity
(with a simple, connected shape, before being torn apart by the turbulent 
flow),
the compression experienced by the set of particles leads to the overall
contraction of
the cloud, given by: 
\begin{equation}
\label{eq:contract_simpl}
	 2 \ln\left(\frac{r}{r_0}\right) \propto  -\Phi_c\frac{t}{t_c}.
\end{equation}
The scalar blob retains its identity for a time $\approx t_c$, so the
particle cloud keeps contracting for a time  $\approx t_c$.
During this time interval, 
the value of $\mathcal{F}$ contracts by an amount, which increases with
$\Phi_c$. \Eref{eq:contract_simpl} suggests a simple relation between
$\mathcal{F}_{\rm min}$ and $\Phi_\mathcal{C}$:
\begin{equation}\label{eq:fmin}
\mathcal{F}_{\rm min} \sim \widetilde{\mathcal{F}}_0\exp(-\gamma \Phi_c),
\end{equation}
where $\gamma$ is a constant {\it a-priori} of order $1$.
We emphasize that this prediction results from the competition between 
the contraction introduced by the scalar, and the mixing by the turbulent flow.

We now turn to DNS runs to test the validity of this phenomenological prediction.
In \fref{fig:r2ncomp} (b) we plot $\mathcal{F}_{\rm min}$ versus $\Phi_{\mathcal{C}}$
on log-linear scale for different particle cloud sizes and Reynolds numbers.
We observe that for all the cases $\mathcal{F}_{\rm min}$ decreases exponentially with
increasing $\Phi_{\mathcal{C}}$, as confirmed by straight lines on log-linear axes and
whose slopes $\gamma$ agree with each other within $15\%$. The inset of 
\fref{fig:r2ncomp} (b) shows the plot of $\tau_{\rm min}/t_c$ versus $\Phi_{\mathcal{C}}$
for the cases considered above. We notice that the time it takes to achieve maximum
contraction of the particle cloud or the maximally de-mixed state for the phoretic
particles has a tendency to saturate as $\Phi_{\mathcal{C}}$ increases. Moreover, we find
that the particle cloud survives only for a fraction of time $t_c$. Our error
bars are large, indicating a need for better statistical averaging; however, we believe 
that our conclusions will not change qualitatively.\\


\begin{figure}
\begin{center}
\includegraphics[scale=0.3]{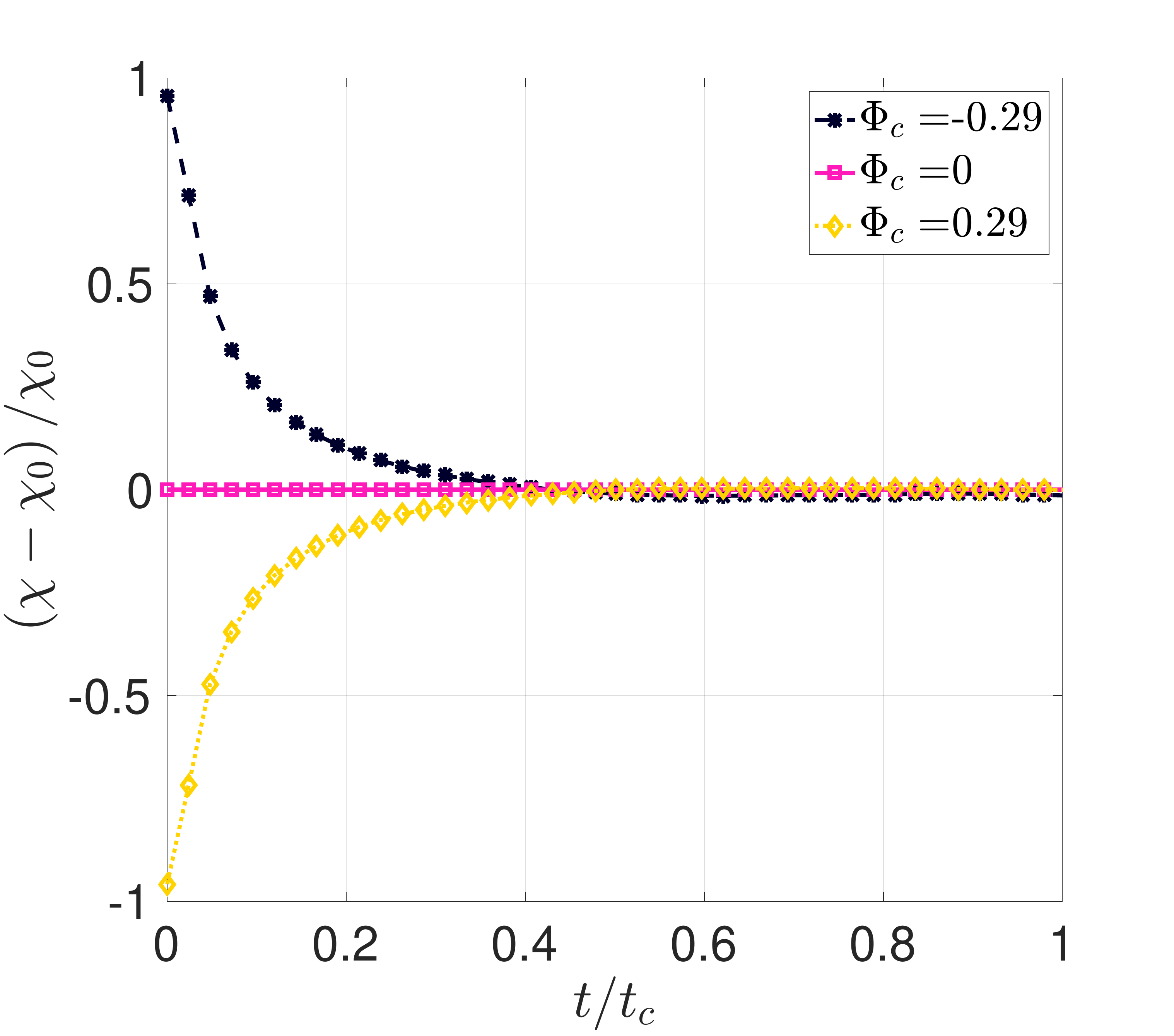}
\put(-125,35){\includegraphics[scale=0.11]{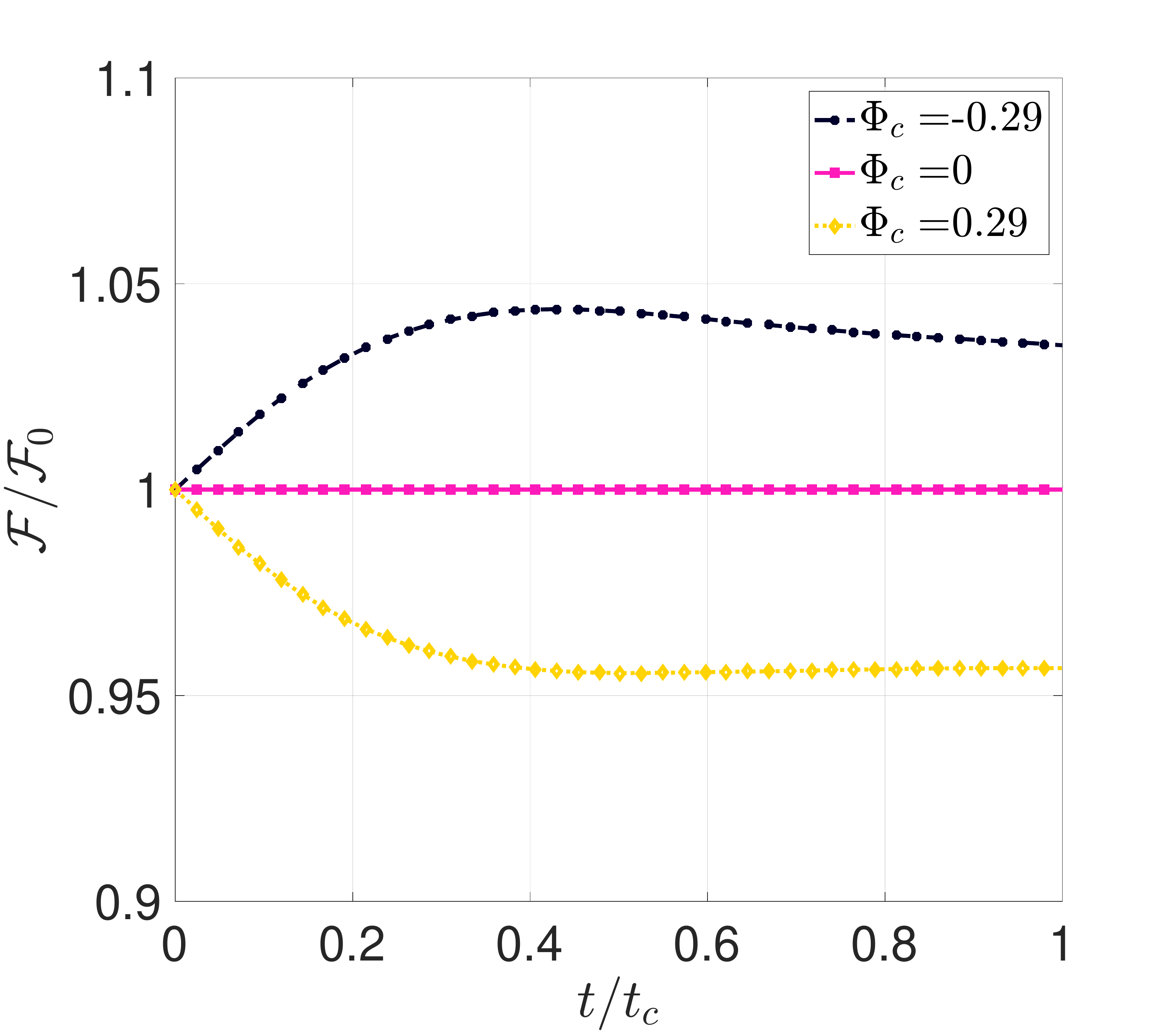}}
\end{center}
\caption{Dilatation factor $\chi=\frac{1}{\cal{F}}\frac{\textrm{d}\cal{F}}{\textrm{d}t}$ and 
in the inset: Mean square separation in presence of phoresis normalized by the non-phoretic case;
from the DNS run with $R_{\lambda}=95$ and $\ell_c/\eta=20.8$.}
\label{fig:FF0}
\end{figure}

As stressed in the introduction, the effect of phoresis 
on the transport properties of the particles 
can be best understood in terms of an effective compressibility. One
way to characterize this compressibility, appropriate in the context of
pair dispersion, is to introduce
the dilation factor : 
\begin{equation}\label{eq:chi}
\chi = \frac{1}{r^2}\frac{\textrm{d}r^2}{\textrm{d}t}=
\frac{1}{\cal F}\frac{\textrm{d}\cal F}{\textrm{d}t}.
\end{equation}
Figure~\ref{fig:FF0} shows the dilation factor $\chi$ estimated 
for the smallest values of $\Phi_{\mathcal{C}}$
represented in \fref{fig:r2ncomp}; as it was difficult to 
see any measurable effect of phoresis in the latter.
Figure~\ref{fig:FF0} indicates that for values of $\Phi_{\mathcal{C}} \approx 0.3$,
the difference between $\chi$ and the corresponding value in the
case of a passive tracer, $\chi_0$, significantly deviates from $0$.
For completeness, the values of $\chi$ for all the values of 
$\Phi_\mathcal{C}$ in \fref{fig:r2ncomp} are shown in \fref{fig:FF0_ext}, see
\ref{appendix:supmat}. It can be seen that even 
for values of $\Phi_\mathcal{C} ={\cal O}(10^{-1})$ 
the impact of phoresis on the initial dilation factor is of order one 
compared to the non-phoretic situation. The inset of \fref{fig:FF0}, 
representing the ratio ${\cal F}/{\cal F}_0$ then shows that although 
the dilation factor relaxes to the non-phoretic value in a timescale 
of a fraction of $t_c$, a sustained modification of the mean square 
pair separation of the order of 5\% compared to the non-phoretic case 
does persist at timescales $t\approx t_c$. It can also be noted that 
these alternative ways of presenting the data better emphasize the  
symmetry between the compressing situation ($\Phi_c>0$, showing a
minimum of ${\cal F}/{\cal F}_0$, which was discussed earlier) 
and the dilating situation ($\Phi_c<0$), showing an equivalent 
maximum of ${\cal F}/{\cal F}_0$.
Thus, a value of $\Phi_c$ of order $10^{-1}$ is sufficient to affect
the rate of dilation of pairs of particles by a significant amount,
albeit over a short time lapse. Yet, as suggested below, these 
effects may be prevalent in conditions where the scalar and the particles
are in a steady state.

\begin{figure}[tb]
\begin{center}
\includegraphics[scale=0.45]{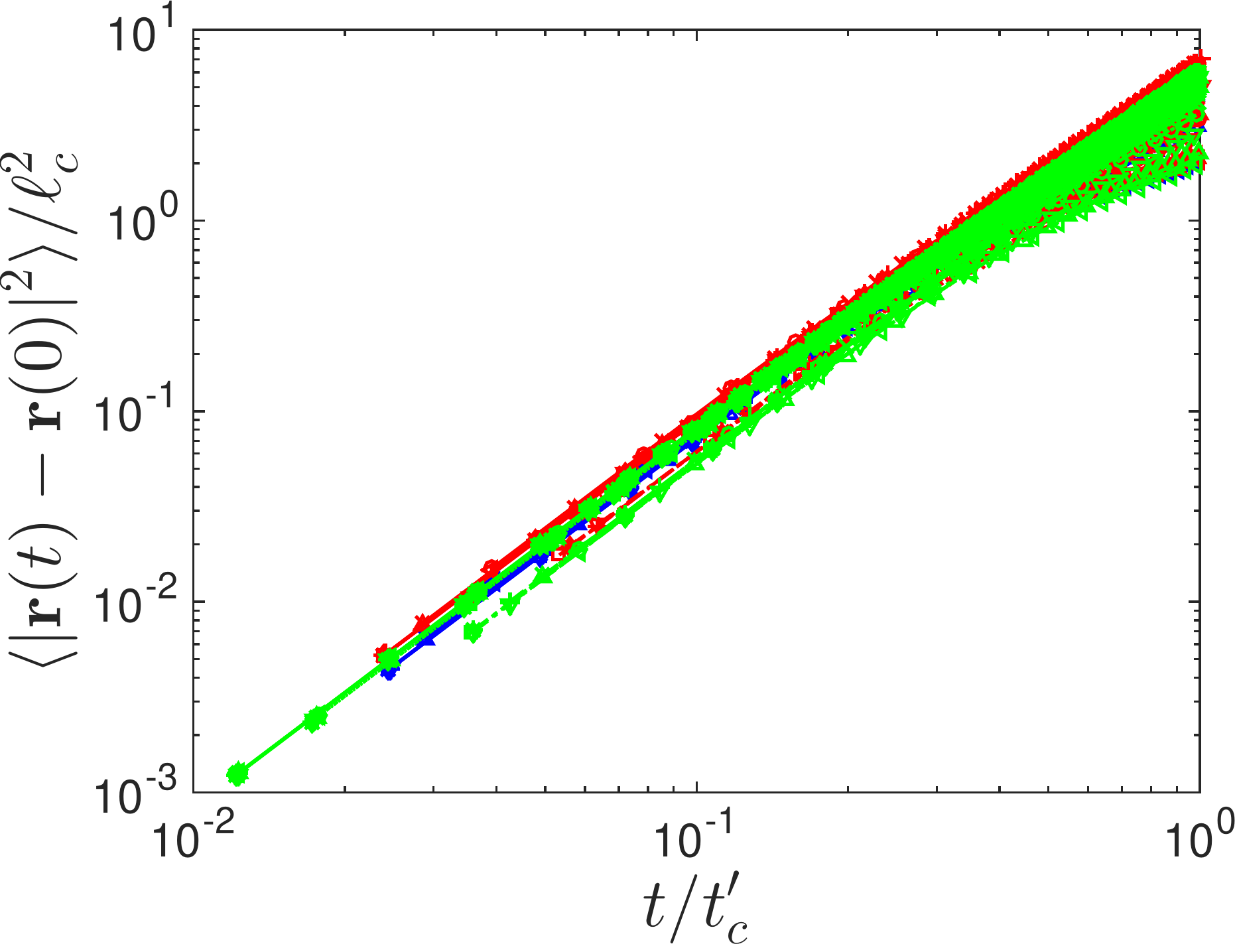}
\end{center}
\caption{
\textbf{Ballistic mean squared relative displacement at short times}
Loglog plots of the mean-squared relative displacement 
$\langle|\delta \mathbf{X}(t)-\delta \mathbf{X}(0)|^2\rangle/\ell^2_c$ versus scaled time $t/t'_c$,
where $t'_{c}=t_c(1+\beta\,\Phi^2_\mathcal{C})^{-\frac{1}{2}}$,
obtained from DNS runs for different values of $\Phi_{\mathcal{C}}$ 
(represented by different symbols) for three different values of Taylor-microscale
Reynolds number $Re_{\lambda}=58$ (blue dashed lines $\ell_c/\eta=10.7$), 
$95$ (red lines: dashed $\ell_c/\eta=20.8$ and colon (:) $\ell_c/\eta=12.0$) and 
$175$ (green lines: dashed $\ell_c/\eta=50.5$, colon (:) $\ell_c/\eta=30.3$ and
dashed-dot $\ell_c/\eta=10.1$); 
different values of particle cloud size $\ell_c$ is indicated by different line-types.
In our simulations $\Delta\theta=1$ and $t_c$ is time scale associated with $\ell_c$.
See Table~\ref{table:Phic} for more details.
}
\label{fig:2partdisp}
\end{figure}

\subsection{Short time separation of the particles in the cloud}

To further understand the mixing and de-mixing process, we look at the two-particle 
statistics, where we compute how the relative separation 
between the particles changes with time. We again turn to phenomenological arguments
to predict the behavior of the mean-squared relative displacement at short-times
\begin{eqnarray}
\langle|\delta \mathbf{X}(t)-\delta \mathbf{X}(0)|^2\rangle 
&\simeq \langle (\delta \mathbf{v})^2 \rangle t^2 \nonumber \\
&\simeq \left[\langle (\delta \mathbf{u})^2 \rangle + \alpha^2 \langle(\delta(\nabla \theta))^2\rangle \right]t^2 \nonumber \\
&\simeq \frac{\ell^2_c}{t^2_c}(1 + \beta \Phi^2_\mathcal{C})t^2,
\label{eq:del_X_sq}
\end{eqnarray}
In going from the first to the second line in the above equation, 
we have neglected the cross term 
$\langle \delta \mathbf{u} \cdot \delta (\nabla \theta) \rangle$. 
This is justified by the complete lack of correlation between the 
injected scalar field $\theta$ and the flow velocity $\mathbf{u}$.
The dimensionless parameter $\beta$ introduced in Eq.~(\ref{eq:del_X_sq}), is
expected to be independent of $\ell_c$ and $Re_\lambda$.
Therefore, we can write
\begin{equation}
	\langle|\delta \mathbf{X}(t)-\delta \mathbf{X}(0)|^2\rangle/\ell^2_c = A [t/t'_{c}]^2,
\end{equation}
with $t'_c=t_c(1+\beta\,\Phi^2_\mathcal{C})^{-\frac{1}{2}}$ and $A$ is a constant.
In \fref{fig:2partdisp} we show the log-log plots of mean-squared relative
displacement $\langle|\delta \mathbf{X}(t)-\delta \mathbf{X}(0)|^2\rangle/\ell^2_c$ versus
scaled time $t/t'_{c}$ for different values of $\Phi_{\mathcal{C}}$ (indicated by
different symbols) for three different Reynolds numbers $Re_{\lambda}=58$ (blue lines),
$Re_{\lambda}=95$ (red lines), and $Re_{\lambda}=175$ (green lines);
and for different particle cloud sizes, see the caption of \fref{fig:2partdisp}, 
as well as 
Table~\ref{table:param} and Table~\ref{table:Phic} in~\ref{appendix:supmat}
for further details.
These DNS runs confirm the phenomenological 
prediction that, when rescaled by the cloud size $\ell_c$, the mean-squared relative displacement is a linear function of $(t/t'_c)^2$ in the ballistic regime.
Moreover, we find that this statistical property is universal, as the plots
for different $\Phi_{\mathcal{C}}$, $Re_{\lambda}$ and $\ell_c$ collapse onto 
each for $\beta\simeq 0.05$.\\

We end this section with a remark that in the present study we are 
interested in an elementary question, whether phoretic effects can lead to 
measurable consequences. For this reason, we focus on a simple quantity,
namely the mean-square separation. However,
much insight can be gained by systematically investigating 
the full distribution of separation at different times. Such an investigation
will be the subject of future work.

\section{Discussion of the orders of magnitude: are phoretic effects observable ?}

Whether phoretic effects (thermo- and diffusio-phoresis, in particular) can
give rise to measurable effects is the important question that we address now. 
The  present discussion is based on the parameter 
$\Phi_c$, introduced in \eref{eq:phic}, which has been demonstrated
in simple problems, see Section~\ref{sec:results},
to provide a good way to quantify the importance of phoretic phenomena:
The larger $\Phi_{\mathcal{C}}$ the larger the effect of phoresis.

\subsection{Revisiting the definition of $\Phi_\mathcal{C}$}
\label{subsec:def_Phi_C}

In terms of possible applications, two effects need to be taken into
account when defining the parameter $\Phi_\mathcal{C}$, relevant to evaluate
the importance of phoretic effects. \\
First, the phoretic
coefficient
$\alpha$ is generally found to depend on the local scalar 
field $\theta$, with a dependence of the form
$\alpha = D_p / \theta$, where $D_p$ has the dimension of a diffusion
coefficient~\cite{bib:NatAbecassis2008,abecassis2009NJP,Vigolo2010}. 
This implies that the dimensionless expression of the flow compressibility,
Eq.~(\ref{eq:phic}) simplifies to:
\begin{equation} \label{eq:phi_tlim_lc}
	\Phi_{\mathcal{C}}=D_p \frac{t_{lim}}{\ell_c^2},
\end{equation}
where $\ell_c$ is the small length scale of the scalar $\theta$
($\nabla^2 \theta \propto \theta/\ell_c^2$), and $t_{lim}$ is the persistence
time of the fluctuations of $\theta$ inducing phoresis.
In our problem, turbulence disrupts a blob of scalar of 
size $\ell_c$, in a time $t_{lim}$ that depends on the $\ell_c$.

Second, whereas in studying phoretic effects in Section~\ref{sec:results},
we
have explicitly considered scalars with a diffusion coefficient of the
order of viscosity: $Pr \approx 1$, it should be kept in mind that 
in practice, the diffusion
coefficient of the scalar can be significantly smaller than viscosity:
$Pr \approx 6$ in the case of thermophoresis, and $Pr \approx 
1000$ in the case of diffusiophoresis. 
We begin this section by discussing the parameter $\Phi_\mathcal{C}$
by taking into account the expression $\alpha = D_p/\theta$, and the large
values of $Pr$.

The following discussion rests on the classical picture of the mixing
of a scalar field by a turbulent flow. 
The fluctuations of the velocity field typically extend over a range of 
length scales,
from the large (stirring) length, $L$, down to the Kolmogorov scale, $\eta$.
A scalar field, advected by the flow, is also subject to molecular diffusion, 
$\kappa  \equiv \nu/Pr$,
where $Pr $ is the Prandtl number already introduced.
In applications with large Prandtl number the scalar fluctuations can extend down 
to scales much smaller than $\eta$. In fact, the smallest scale of the
scalar fluctuations, $\eta_B$, known as the Batchelor scale, is of order
$\eta_B\equiv \eta Pr^{-1/2}$ for $Pr\gg 1$~\cite{Batchelor:59}. \\
The persistence of scalar fluctuations at a scale $\ell_c$ depends on
the range of scale. Namely, when $\eta \le \ell_c \le L$ (the size of the
scalar blob lies in the inertial range), the blob is subject to the turbulent
shear, $\propto (\varepsilon/\ell_c^2)^{1/3}$, so the persistent time is
$t_{lim} \propto (\ell_c^2/\varepsilon)^{1/3}$. This is the case we have
considered so far. In this case, the expression~(\ref{eq:phi_tlim_lc})
reduces to 
\begin{equation}\label{eq:phi_eta_lc_nu}
	\Phi_{\mathcal{C}} = \frac{D_p}{\nu}\left(\frac{\eta}{\ell_c}\right)^{4/3}.
\end{equation}
which immediately shows that the largest possible value of $\Phi_\mathcal{C}$
is obtained when $\ell_c \approx \eta$: $\Phi_\mathcal{C} \lesssim D_p/\nu$.
In the other regime, $\eta_B \le \ell_c \le \eta$, the velocity field
is smooth at the scale $\ell_c$, and the strain acting on a blob of size
$\ell_c$ is $\propto (\varepsilon/\nu)^{1/2}$. This implies that the time
$t_{lim} \propto (\nu/\varepsilon)^{1/2}$, which is also known as the
Kolmogorov time scale~\cite{Frisch}. In this case, the expression for 
$\Phi_\mathcal{C}$ reduces to:
\begin{equation}\label{eq:phi_etaB_lc}
	\Phi_{\mathcal{C}} = \frac{D_p}{\ell_c^2}\times\left(\frac{\nu}{\varepsilon}\right)^{1/2}  = \frac{D_p}{\nu} \times \left( \frac{\eta}{\ell_c} \right)^2
= \frac{D_p}{\kappa} \times \left( \frac{\eta_B}{\ell_c} \right)^2
\end{equation}
Eq.~\ref{eq:phi_etaB_lc} shows that the value of $\Phi_\mathcal{C}$ can vastly
exceed $D_p/\nu$, the limiting value when $\ell_c \ge \eta$. 
Specifically, in the case of blobs of size 
$\ell_c \approx \eta_B$, 
$\Phi_{\mathcal{C}} \approx D_p/\kappa$.   
These expressions are particularly relevant, given 
the large values of $Pr$ in the relevant cases of thermo- and 
diffusiophoresis.

To summarize, the estimates given so far show that the maximum strength of the 
phoresis effects quantified in terms of the parameter $\Phi_{\mathcal{C}}$
can be ultimately expressed 
as ratio of the diffusion coefficient $D_p$ and either the viscosity $\nu$ or
the diffusion coefficient of the scalar field (temperature or salt) $\kappa$.


\subsection{How large are phoretic effects ? }
\label{subsec:how_large}

To proceed, we use the values of $D_p$ reported in the literature.
Microfluidic 
experiments have shown so far a maximum effect when using carboxylate 
colloids in gradients of Lithium Chloride (LiCl), this combination 
of particle and salt lead to a diffusiophoretic coefficient as 
large as $D_p\approx 300\mu$m$^2$s$^{-1}$. 
In experiments on thermophoresis values of $D_p$ as 
large as $3000\mu$m$^2$s$^{-1}$ have been measured~\cite{Vigolo2010},
hence 10 times larger than in the diffusiophoresis case.
These estimates are consistent with those measured in the 
oceans~\cite{thorpe}.

This implies that the ratio $D_p/\nu$, which corresponds to the largest
possible value of $\Phi_\mathcal{C}$ when $\ell_c$ is in the inertial range,
see~\eref{eq:phi_eta_lc_nu},
is of the order $10^{-4}$ in the case of diffusiophoresis, and of
order $3 \times 10^{-3}$ in the case of thermophoresis.

These values are clearly smaller than the values necessary to
measure any of the effects discussed in Section~\ref{sec:results}.
Assuming, however, a value of $\ell_c $ smaller than $\eta$, but larger 
than $\eta_B$
shows that the values of $\Phi_\mathcal{C}$ cannot be larger than
$\Phi_\mathcal{C} \lesssim 0.02$ in the case of thermophoresis, 
and $\Phi_\mathcal{C} \lesssim 0.1$ in the case of diffusiophoresis.
These values are sufficient to observe, in principle, a significant 
effect of diffusiophoresis.

It is useful to compare the previous estimates with existing experimental 
results, where effects of phoresis were unambiguously found.

In the microfluidic
experiments on diffusiophoresis by 
Abecassis \emph{et al.}~\cite{bib:NatAbecassis2008,abecassis2009NJP}, 
using a laminar flow,
the limiting time is given by the mean advection $t_{lim} = z_{obs} / U_0$ 
(where $z_{obs}$ is the observation position along the micro-channel 
and $U_0$ is the advection velocity along the channel. Interestingly, 
$\Phi_{\mathcal{C}}$ in their experiment can then be estimated to be in the range 
$[10^{-3}-10^{-1}]$, with measurable effects on the spreading and 
focusing of the concentration profile of an initial colloidal band.

For the experiments in chaotic flows by 
Desseigne \emph{et al.}~\cite{deseigne2014softmat} and 
Mauger \emph{et al.}~\cite{Mauger16}, which are closer in spirit to
the present work, the limiting time scale at the 
early stage of the process is given by the period of the chaotic cycle 
(at larger times the diffusion time of salt also becomes important), 
leading to typical values of $\Phi_{\mathcal{C}}$ of the order of $[10^{-3}-10^{-2}]$, 
with subtle but still measurable effects on the concentration field of 
colloids and its gradients. 
We also notice that Schmidt \emph{et al.}~\cite{Schmidt2016} report a 
possible observation of diffusiophoresis, by looking at clustering properties 
of particles at inertial scales in a turbulent gravity flow with salt 
gradients. In such conditions, the largest 
expected value for $\Phi_{\mathcal{C}}$ is of the order of $D_p/\nu$ (with $\nu$ 
the kinematic viscosity of the fluid), which in their experiment is of the 
order of $10^{-4}$. 

It is worth pointing out that in the problem of pair dispersion, 
as we have studied it numerically, the initial scalar distribution affects 
colloid transport for a small time only. Scalar gradients, in a 
statistically steady-state regime, may act cumulatively, resulting
in much larger effects.
As such, the requirement in term of $\Phi_{\mathcal{C}}$ 
to observe a significant phoretic effect could be conceivably much reduced,
compared to what we found in the purely transient problem investigated here.

The above discussion and the values of $\Phi_{\mathcal{C}}$ obtained on the basis
of~\cite{Vigolo2010,thorpe} indicate that the phoretic effects can be important 
and therefore may affect the transport
of small organisms, or pollutant particles, as a result of local
gradients of salinity, dissolved oxygen, temperature, etc.

\section{Summary}
In this work, we have explored the interplay 
between turbulent transport and
phoretic effects. The specific problem investigated is purely transient, and
concerns a cloud of colloids, released with a blob of scalar of size
$\ell_c$ in an otherwise statistically stationary flow.

The phenomena discussed here results from a competition between the
compressibility of the 
particle velocity field,~\eref{eq:partvel}, and the fast 
dispersion of the scalar blob. On dimensional grounds,
this competition can be described at short times by the 
dimensionless ratio $\Phi_\mathcal{C}$, defined by \eref{eq:phic}.

In the case where the scalar generates an effective compressibility 
($\Phi_\mathcal{C} > 0$), we observed that particles tend to come together,
over a time which is a fraction of the time $t_c $, characteristic
of the size $\ell_c$. The minimum in the mutual distance between particles can
be approximated as
a function of $\Phi_\mathcal{C}$, which decays exponentially at large
values of $\Phi_\mathcal{C}$. The quantity $\Phi_\mathcal{C}$ provides a satisfactory
way to describe the initial stage of the separation between particles.

We have used our numerical results, as well as those from previous studies, to
discuss the significance of the approach followed in the present study for the
existing or possible experiments and the natural flows.
The discussion of the previous section indicates that in 
order to maximize the chances to observe a signature of 
diffusiophoresis at inertial scales of turbulence, one has to: 
(i) use an appropriate indicator (e.g., the dilation factor for pair 
separation diagnosis) and (ii) maximize $D_p$ and minimize both 
$\epsilon$ and the observation scale $\ell_c$.

We conclude by recalling that the present study has been focused on 
the transient problem, where the scalar field is injected together with 
the particles. 

In this time-dependent problem, 
the colloid velocity field is either manifestly attracting or 
repelling, possibly leading to the strong mixing or de-mixing effects,
illustrated in figures~\ref{fig:partcloud} and~\ref{fig:r2ncomp}. 
Such strong effects are not
expected when colloids are injected in a scalar field in a statistically 
steady state, as the divergence of the colloid velocity field
$\nabla \cdot \mathbf{v}$, is equally likely to be positive or negative. 
The related but distinct problem of the properties of the colloid distribution
in the presence of a of turbulent scalar field in a steady state
therefore deserves further attention. 
In fact, as it is the case in the problem of inertial particles in
a turbulent flow~\cite{Max87,BFF01,FalkPum04}, the compressibility of the velocity 
field experienced by the particles is likely to lead to preferential 
concentration. It will be interesting to explore the analogies and
differences between the two problems, in particular in terms of particle
dispersion properties~\cite{Bec+10,Bragg+16}.

\bigskip
\bigskip
\noindent
{\bf Acknowledgements:}
We are very thankful to C. Cottin-Bizonne, F. Raynal and C. Ybert 
for their insight into the physics of phoretic mechanisms. 
V.S. also thanks D. Buaria for useful discussions. 
This work was supported by the European project EuHIT - European 
High-performance Infrastructures in
Turbulence, and by French research programs ANR-16-CE30-0028 and 
LABEX iMUST (ANR-10-LABX-0064) of Universit\'e de Lyon, within the 
program``Investissements d'Avenir" (ANR-11-IDEX-0007).

\appendix
\section{Supplementary information on the DNS runs}
\label{appendix:supmat}

This appendix presents a list of the flows we have simulated (Table~\ref{table:param}), and of the values of $\alpha$ and $\Phi_{\cal{C}}$ used
(Table~\ref{table:Phic}).

\begin{figure*}
\begin{center}
\includegraphics[scale=0.45]{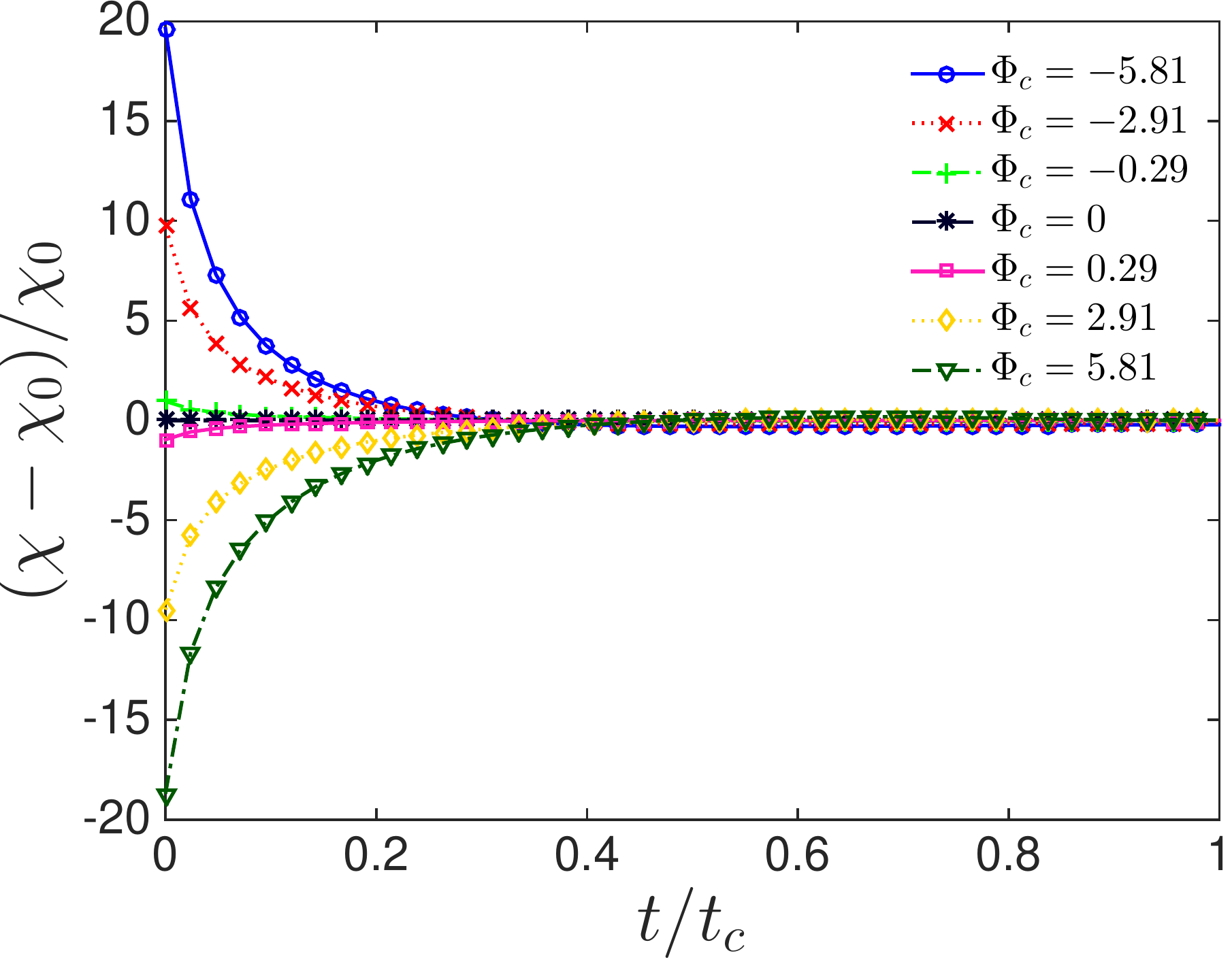}
\put(-180,160){\large{\bf(a)}}
\includegraphics[scale=0.45]{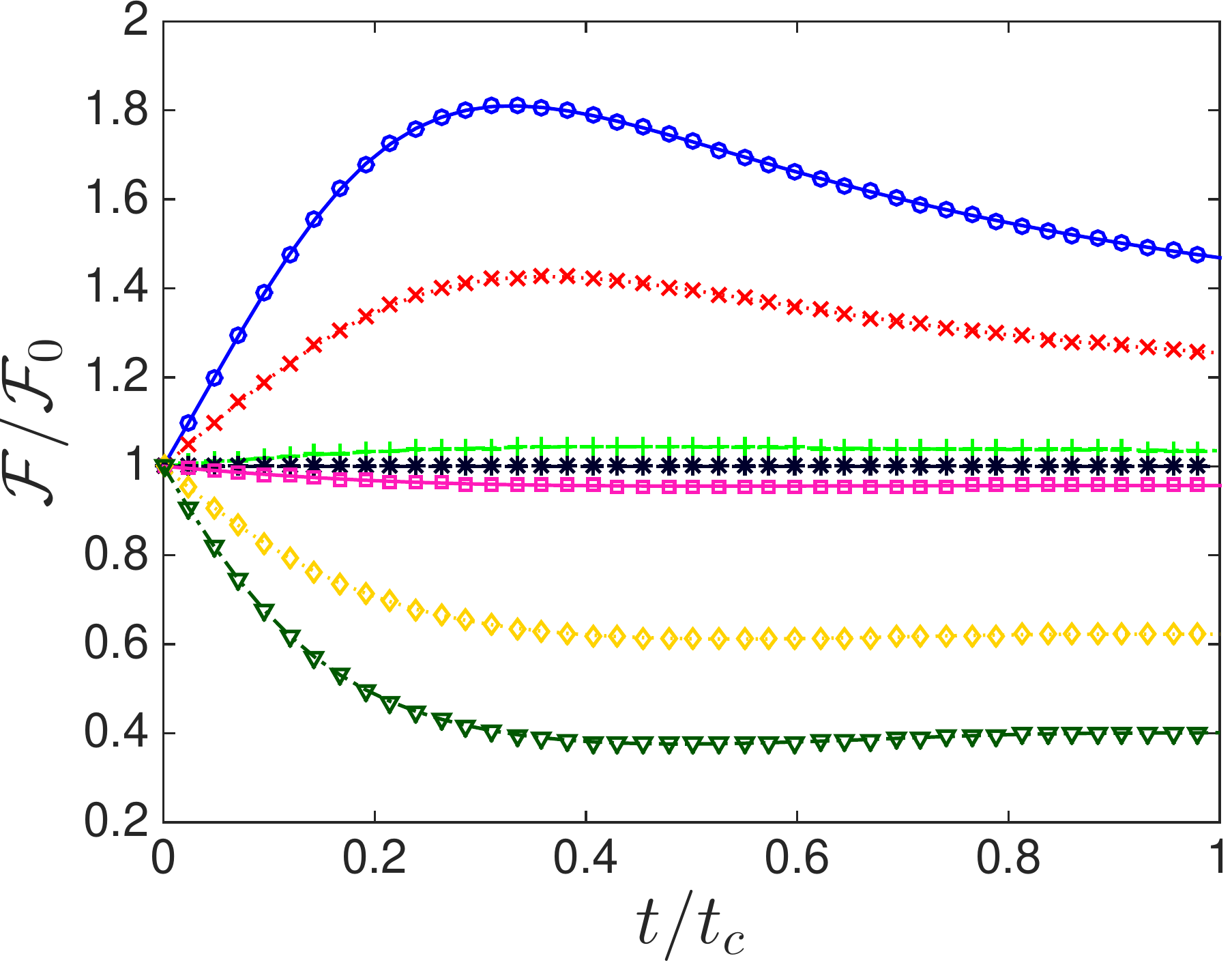}
\put(-180,160){\large{\bf(b)}}
\end{center}
\caption{(a) Dilatation factor $\chi=\frac{1}{\cal{F}}\frac{\textrm{d}\cal{F}}{\textrm{d}t}$ 
and (b) mean square separation in presence of phoresis normalized by the non-phoretic case,
from the DNS run with $R_{\lambda}=95$ and $\ell_c/\eta=20.8$.}
\label{fig:FF0_ext}
\end{figure*}

In addition, \fref{fig:FF0_ext} presents the values of $\chi$ and of the
mean separation ${\cal F}/{\cal F}_0$ for values of $\Phi_{\cal C}$ larger
than those shown in \fref{fig:FF0} of the main text.

\begin{table*}
\begin{center}
   \begin{tabular}{@{\extracolsep{\fill}} c c c c c c c c c c c c c}
    \hline
    $ $ & $N_c$ & $\nu\times 10^{-3}$ & $\kappa\times 10^{-3}$ & 
    $R_{\lambda}$ & $u_{rms}$ & 
    $L_0$ & $\lambda$ & $\eta$ & $k_{max}\eta$ & 
    $\epsilon$ & $\ell_{c}/\eta$ & $\beta$\\
   \hline \hline

   {\tt Run1} & $96$ & $7.5$ & $15.0$ & $58$ & $0.728$ &
   $1.38$ & $0.596$ & $0.04$ & $1.79$ &
   $0.168$ & $10.7$ & $0.05$\\
   {\tt Run2} & $192$ & $3.0$ & $6.0$ & $95$ & $0.7289$ &
   $1.30$ & $0.39$ & $0.0204$ & $1.83$ &
   $0.157$ & $20.8$ & $0.05$\\
   {\tt Run3} & $192$ & $3.0$ & $6.0$ & $95$ & $0.7289$ &
   $1.30$ & $0.39$ & $0.0204$ & $1.83$ &
   $0.157$ & $12.0$ & $0.05$\\
   {\tt Run4} & $384$ & $0.94$ & $1.88$ & $175$ & $0.75$ &
   $1.23$ & $0.22$ & $0.0084$ & $1.52$ & 
   $0.168$ & $50.5$ & $0.05$\\
   {\tt Run5} & $384$ & $0.94$ & $1.88$ & $175$ & $0.75$ &
   $1.23$ & $0.22$ & $0.0084$ & $1.52$ & 
   $0.168$ & $30.3$ & $0.06$\\
   {\tt Run6} & $384$ & $0.94$ & $1.88$ & $175$ & $0.75$ &
   $1.23$ & $0.22$ & $0.0084$ & $1.52$ & 
   $0.168$ & $10.1$ & $0.07$\\
\hline
\end{tabular}
\end{center}
\caption{\small 
Parameters for the DNS runs $\tt Run1-Run6$: 
$N^3_c$ the number of collocation points, $\nu$ the viscosity,
$\kappa$ the diffusivity, $R_{\lambda}=u_{\rm rms}\lambda/\nu$ 
the Taylor-scale based Reynolds number, $u_{\rm rms}=\sqrt{2\,E/3}$ the root-mean-squared
velocity and $\lambda=u_{\rm rms}/\sqrt{\langle(\partial_{\rm x}u)^2\rangle}$ the Taylor-microscale,
$L_0=(\pi/2u^2_{\rm rms})\sum_{\rm k} E(k)/k$ the integral length scale, 
$\eta=(\nu^2/2\Omega)^{1/4}$ the Kolmogorov dissipation length scale, 
$\epsilon$ the energy injection rate, and $\ell_c$ the measure of size of the Gaussian scalar blob. 
$E$ is the mean kinetic energy and
$\langle(\partial_{\rm x}u)^2\rangle=2\Omega/15$, where $\Omega$ is the enstrophy. 
In our DNSs we have fixed the Prandtl number $Pr=\nu/\kappa$ at $0.5$. 
}
\label{table:param}
\end{table*}

\begin{table*}
\begin{center}
   \begin{tabular}{@{\extracolsep{\fill}} c c c c c | c c c c}
\hline
$$ & $$ & $\Phi_{\mathcal{C}}$ & $$ & $$ & $$ & $\Phi_{\mathcal{C}}$ & $$\\   
\hline    
$Symbol$ & $\alpha$ & $\tt Run1$ & $\tt Run2$ & $\tt Run3$ & $\alpha$ & $\tt Run4$ & $\tt Run5$ & $\tt Run6$\\

   \hline \hline
   $\bigcirc$ & $-1.0$  &  $-5.69$ &  $-5.81$ & $-12.10$ &
   $-0.1$ & $-0.569$  & $-1.12$ & $-4.86$\\

   $X$ & $-0.5$  &  $-2.84$ &  $-2.91$ & $-6.05$ &
   $-0.075$ & $-0.43$  & $-0.84$ & $-3.65$\\

   $+$ & $-0.1$  &  $-0.57$ &  $-0.58$ & $-1.21$  &
   $-0.05$ & $-0.28$  & $-0.56$ & $-2.43$\\

   $\ast$ & $-0.05$  &  $-0.28$ &  $-0.29$ & $-0.60$ &
   $-0.01$ & $-0.057$ & $-0.11$ & $-0.47$\\

   $\square$ & $0$ &   $0$ & $0$ & $0$ & $0$ & $0$ & $0$ & $0$\\
   
   $\diamondsuit$ & $0.05$ &  $0.28$ &  $0.29$ & $0.60$  &
   $0.01$ & $0.057$  &  $0.11$  &  $0.49$\\

   $\nabla$ & $0.1$ &  $0.57$ &  $0.58$ & $1.21$  &  
   $0.05$ & $0.28$ &  $0.56$ &  $2.43$\\

   $\triangle$ & $0.5$ &  $2.84$ &  $2.91$ & $6.05$ & 
   $0.075$ & $0.43$  &  $0.84$ &  $3.65$\\

   $\triangleleft$ & $1.0$ & $5.69$  &  $5.81$ & $12.10$  &
   $0.1$ & $0.57$  & $1.12$ & $4.86$\\

\hline
\end{tabular}
\end{center}
\caption{\small Values of the dimensionless number $\Phi_{\mathcal{C}}$ from our
	DNS runs $\tt Run1$-$\tt Run6$ for nine different values of $\alpha$.
	The symbols in the first column serve as guide for the curves in \fref{fig:2partdisp}.
}
\label{table:Phic}
\end{table*}

\bibliographystyle{iopart-num}

\section*{References}


\begin{thebibliography}{10}
\expandafter\ifx\csname url\endcsname\relax
  \def\url#1{{\tt #1}}\fi
\expandafter\ifx\csname urlprefix\endcsname\relax\def\urlprefix{URL }\fi
\providecommand{\eprint}[2][]{\url{#2}}

\bibitem{anderson1989colloid}
Anderson J~L 1989 {\em Ann. Rev. Fluid Mech.\/} {\bf 21} 61--99

\bibitem{li2008encyclopedia}
Li D 2008 {\em Encyclopedia of microfluidics and nanofluidics\/} (Springer
  Science \& Business Media)

\bibitem{zborowski2015magnetophoresis}
Zborowski M and Chalmers J~J 2015 {\em Wiley Encyclopedia of Electrical and
  Electronics Engineering\/}

\bibitem{bib:NatAbecassis2008}
Ab{\'e}cassis B, Cottin-Bizonne C, Ybert C, Ajdari A and Bocquet L 2008 {\em
  Nature materials\/} {\bf 7} 785--789

\bibitem{abecassis2009NJP}
Abecassis B, Cottin-Bizonne C, Ybert C, Ajdari A and Bocquet L 2009 {\em New J.
  Phys.\/} {\bf 11} 075022

\bibitem{deseigne2014softmat}
Deseigne J, Cottin-Bizonne C, Stroock A~D, Bocquet L and Ybert C 2014 {\em Soft
  matter\/} {\bf 10} 4795--4799

\bibitem{PREVolk2014}
Volk R, Mauger C, Bourgoin M, Cottin-Bizonne C, Ybert C and Raynal F 2014 {\em
  Physical Review E\/} {\bf 90} 013027

\bibitem{Mauger16}
Mauger C, Volk R, Machicoane N, Bourgoin M, Cottin-Bizonne C, Ybert C and
  Raynal F 2016 {\em Phys. Rev. Fluids\/} {\bf 1}(3) 034001

\bibitem{Shaw03}
Shaw R 2003 {\em Ann. Rev. Fluid Mech.\/} {\bf 35} 183--227

\bibitem{BalEaton:2010}
Balachandar S and Eaton J~K 2010 {\em Ann. Rev. Fluid Mech.\/} {\bf 42}
  111--133

\bibitem{MaxRil83}
Maxey M~R and Riley J~J 1983 {\em Phys. Fluids\/} {\bf 26} 883--889

\bibitem{Gatignol83}
Gatignol R 1983 {\em J. M\'ec. Th\'eor. Appl.\/} {\bf 1} 143--160

\bibitem{BFF01}
Balkovsky E, Falkovich G and Fouxon A 2001 {\em Phys. Rev. Lett.\/} {\bf 86}
  2790--2793

\bibitem{FFS02}
Falkovich G, Fouxon A and Stepanov M~G 2002 {\em Nature\/} {\bf 419} 151--154

\bibitem{FalkPum04}
Falkovich G and Pumir A 2004 {\em Phys. Fluids\/} {\bf 16} L47--50

\bibitem{Falkovich2014}
Belan S, Fouxon I and Falkovich G 2014 {\em Phys. Rev. Lett.\/} {\bf 112}(23)
  234502

\bibitem{Schmidt2016}
Schmidt L, Fouxon I, Krug D, van Reeuwijk M and Holzner M 2016 {\em Phys. Rev.
  E\/} {\bf 93}(6) 063110

\bibitem{de2016clustering}
De~Lillo F, Cencini M, Musacchio S and Boffetta G 2016 {\em Physics of
  Fluids\/} {\bf 28} 035104

\bibitem{mitra2016turbophoresis}
Mitra D, Haugen N~E~L and Rogachevskii I 2016 {\em arXiv preprint
  arXiv:1603.00703\/}

\bibitem{Frisch}
Frisch U 1995 {\em Turbulence\/} 1st ed (Cambridge University Press) ISBN 0 521
  45713 0

\bibitem{Pum94}
Pumir A 1994 {\em Phys. Fluids\/} {\bf 6} 2118--2132

\bibitem{Bragg+14}
Bragg A~D and Collins L~R 2014 {\em New J. Phys.\/} {\bf 16} 055013

\bibitem{Vigolo2010}
Vigolo D, Rusconi R, Stone H~A and Piazza R 2010 {\em Soft Matter\/} {\bf 6}
  3489 ISSN 1744-683X

\bibitem{Batchelor:59}
Batchelor G~K 1959 {\em J. Fluid Mech.\/} {\bf 5} 113--133

\bibitem{thorpe}
Thorpe S~A 2005 {\em The turbulent ocean\/} 1st ed (Cambridge University Press)
  ISBN 13 978-0-521-83543-5

\bibitem{Max87}
Maxey M~R 1983 {\em J. Fluid Mech.\/} {\bf 174} 441--465

\bibitem{Bec+10}
Bec J, Biferale L, Lanotte A~S, Scagliarini A and Toschi F 2010 {\em J. Fluid
  Mech.\/} {\bf 645} 497--528

\bibitem{Bragg+16}
Bragg A~D, Ireland P~J and Collins L~R 2016 {\em Phys. Fluids\/} {\bf 28}
  013305

\end{thebibliography}
\providecommand{\newblock}{}

\end{document}